              \documentclass[twocolumn,amsmath,amssymb,10pt,superscriptaddress,a4paper,letterpaper,fleqn]{revtex4-1}
              \usepackage{amssymb}
              \usepackage{epsfig}
              \usepackage{graphicx}
              \usepackage{dcolumn}
              \usepackage{array}
              \usepackage{bm}
              \usepackage{fancyheadings}
              \usepackage{longtable}
              \usepackage{multirow}
              \usepackage{float}
              \usepackage{afterpage}
              \usepackage{color}
               
              \setlongtables
              \usepackage[breaklinks=true,linkbordercolor={1 1 1}]{hyperref}
               
              \usepackage{accents}
              \newcommand{\bsigma}{\boldsymbol\sigma}
               
\begin{document}\setcounter{page}{1}
               
              %%% **********************************************************************
               
              \title{
              %% Please do not remove the line below
              %\qquad \\ \qquad \\ \qquad \\  \qquad \\  \qquad \\ \qquad \\
              \qquad \\ \qquad \\ \qquad \\ 
              Wanted! Nuclear Data for Dark Matter Astrophysics}
               
              \author{P. Gondolo}
              \email[Corresponding author: ]{paolo.gondolo@utah.edu}
              \affiliation{Department of Physics and Astronomy, University of Utah, Salt Lake City, UT 84112-0830, USA}
               
              \date{\today} 
              %\received{8 March 2013; revised received XX June 2013; accepted XX September 2013}
               
              \begin{abstract}
              Astronomical observations from small galaxies to the largest scales in the universe can be consistently
              explained by the simple idea of dark matter. The nature of dark matter is however still unknown. Empirically
              it cannot be any of the known particles, and many theories postulate it as a new elementary particle. Searches
              for dark matter particles are under way: production at high-energy accelerators, direct detection through
              dark matter-nucleus scattering, indirect detection through cosmic rays, gamma rays, or effects on stars.
              Particle dark matter searches rely on observing an excess of events above background, and a lot of controversies
              have arisen over the origin of observed excesses. With the new high-quality cosmic ray measurements from the
              AMS-02 experiment, the major uncertainty in modeling cosmic ray fluxes is in the nuclear physics cross sections
              for spallation and fragmentation of cosmic rays off interstellar hydrogen and helium. The understanding of
              direct detection backgrounds is limited by poor knowledge of cosmic ray activation in detector materials, with
              order of magnitude differences between simulation codes. A scarcity of data on nucleon spin densities blurs
              the connection between dark matter theory and experiments. What is needed, ideally, are more and better
              measurements of spallation cross sections relevant to cosmic rays and cosmogenic activation, and data on the
              nucleon spin densities in nuclei.
              \end{abstract}
              \maketitle
               
              %%% DO NOT EDIT the following section enclosed by *****
              %%% ***************************************************
              \lhead{Wanted! Nuclear Data $\dots$}
              \chead{NUCLEAR DATA SHEETS}
              \rhead{P. Gondolo}
              \lfoot{}
              \rfoot{}
              \renewcommand{\footrulewidth}{0.4pt}
              %%% ***************************************************
               
              \newcommand{\p}{{\rm p}}
              \newcommand{\n}{{\rm n}}
              \newcommand{\ip}[2]{{}^{\rm #2}{{\rm #1}}}
               
              Cosmological observations agree with a universe made mostly of dark energy and cold dark matter. Their nature is
              still unknown. In the quest to unveil what dark matter is, there appears to be a need for new or better nuclear physics data.
               
              This short article starts by overviewing the cold dark matter problem: the issue, the simplest idea of a new elementary
              particle, and some ways to test this idea. This is followed by three uncertain nuclear physics aspects of relevance to
              the dark matter problem, which are at the same time a request for more information: $A(\p,x)B$ and $A(\alpha,x)B$ cross
              sections up to 100 GeV of beam energy for stable and long-lived ($\gtrsim 1$ Myr) isotopes up to $A\sim 64$ (admittedly
              a tall order); $\ip{Ge}{nat}(\n,x)B$, $\ip{Xe}{nat}(\n,x)B$ and $\ip{Ar}{nat}(\n,x)B$ cross sections around 1 MeV of
              beam energy; nucleon spin densities up to $\sim100$ MeV/$c$ of momentum transfer ($\sim2$ fm$^{-1}$) inside $\ip{C}{13}$,
              $\ip{O}{17}$, $\ip{F}{19}$, $\ip{Na}{23}$, $\ip{Ca}{43}$, $\ip{Ge}{73}$, $\ip{I}{127}$, $\ip{Xe}{129,131}$, $\ip{Cs}{133}$,
              $\ip{W}{183}$, which are nuclei used or soon to be used in dark matter experiments.
               
              \vspace{0.5\baselineskip}
              \begin{center}
              {\bf I. ~ THE COLD DARK MATTER PROBLEM}
              \end{center}
               
              Modern cosmology has achieved the measurement in physical units of the energy density of the universe constituents. The
              most precise method is based on applying the well-known atomic physics of hydrogen and helium ionization and recombination,
              plus general relativity, to the universe of 13 billions years ago. A recent analysis~\cite{WMAP9} of several cosmological
              data shows that the universe is composed mostly of dark energy and cold dark matter: $585\pm3$ pJ/m$^3$ in dark energy,
              $194\pm3$ pJ/m$^3$ in cold dark matter, $37.6\pm0.5$ pJ/m$^3$ in ordinary matter, 1 to 7 pJ/m$^3$ in neutrinos, and
              $0.04175\pm0.00004$ pJ/m$^3$ in photons. Here ``matter'' is defined by its equation of state $p\ll\rho$, where $p$ is
              the pressure and $\rho$ the total energy density including rest mass. ``Dark energy'' is defined by its equation of
              state $p=-\rho$ (cosmological constant). Matter is subdivided into ``ordinary matter,'' which in this context is
              protons, neutrons, and electrons, and ``cold dark matter,'' which does not interact significantly with photons and
              ordinary matter at hydrogen recombination.
               
              The amount and location of cold dark matter is inferred from a variety of cosmological data ranging from dwarf galaxies
              to the largest structures in the universe. Galaxies, through rotation curves and velocity dispersion profiles, are
              observed to spin faster or be hotter than the gravity which the visible mass can support. Clusters of galaxies, through
              the motion of galaxies, gravitational lensing, and measurements of the gas density and pressure, are observed to be
              mostly made of invisible mass. The presence of an invisible mass  in the universe of $\sim$13 billion years ago is also
              the simplest way to understand how the inhomogeneities observed in the young universe through the Cosmic Microwave
              Background (CMB) have evolved into the observed distribution of galaxies. 
               
              What is cold dark matter made of? Of course it cannot be photons, but it cannot either be any of the unstable particles
              in the standard model of particle physics, because they do not live for billions of years. Nor can it be protons, neutrons,
              or electrons, because they would have coupled to the CMB photons. Finally it cannot be standard model neutrinos either
              because their mass is too small. No known particle can be cold dark matter: this is the dark matter problem.
               
              The simplest and most elegant idea is that cold dark matter is a new massive elementary particle that interacts weakly, a
              WIMP (for Weakly Interacting Massive Particle). One naturally obtains the right cosmic density of WIMPs, and more
              importantly one can experimentally test the WIMP hypothesis because the same physical processes that produce the
              right density of WIMPs make their detection possible.
               
              The WIMP cosmic density is set by WIMP production and annihilation in the primordial universe, e.g., quark-antiquark
              annihilation into WIMP-antiWIMP, or WIMP-WIMP if an antiWIMP is identical to a WIMP: $q\overline{q} \to \chi
              \accentset{(-)}{\chi}$ and its inverse $ \chi \accentset{(-)}{\chi} \to q \overline{q}$. WIMP production may also
              occur at high energy particle accelerators, which may lead to  the discovery of dark matter in the laboratory. WIMP
              annihilation in galactic halos or astrophysical objects like stars may allow the indirect detection of WIMPs through
              their annihilation signals. The crossed reaction $q\chi \to q\chi$ may allow the direct detection of WIMPs by scattering
              of galactic WIMPs in laboratory detectors. Scattering also sets the size of the smallest dark halos in the universe.
               
              During the years, claims of WIMP detection have come and gone. Among the current claims are (i) the detection of an
              excess microwave emission around the Galactic Center (the WMAP/Planck haze), (ii) an annual modulation in the direct
              detection rate expected from the motion of the Earth around the Sun~\cite{Bernabei1997Aalseth2011}, (iii) a tentative
              detection of gamma-ray lines at $\sim$130 GeV photon energy from regions near the Galactic Center~\cite{Weniger2012},
              and (iv) an excess in the flux of cosmic ray positrons above $\sim$10 GeV~\cite{positrons,Aguilar2013}.
               
              It is the latter evidence that has brought the author to consider the nuclear physics aspects of dark matter searches.
              The excess is claimed over an expected background due to positron production by cosmic ray collisions in the galaxy.
              How well is this background predicted? It turns out that nuclear physics uncertainties are important, as described
              in the next section.
               
              \vspace{0.5\baselineskip}
              \begin{center}
              {\bf II. ~ COSMIC RAY BACKGROUNDS}
              \end{center}
               
              \begin{figure}[!t]
              \includegraphics[width=0.7\columnwidth]{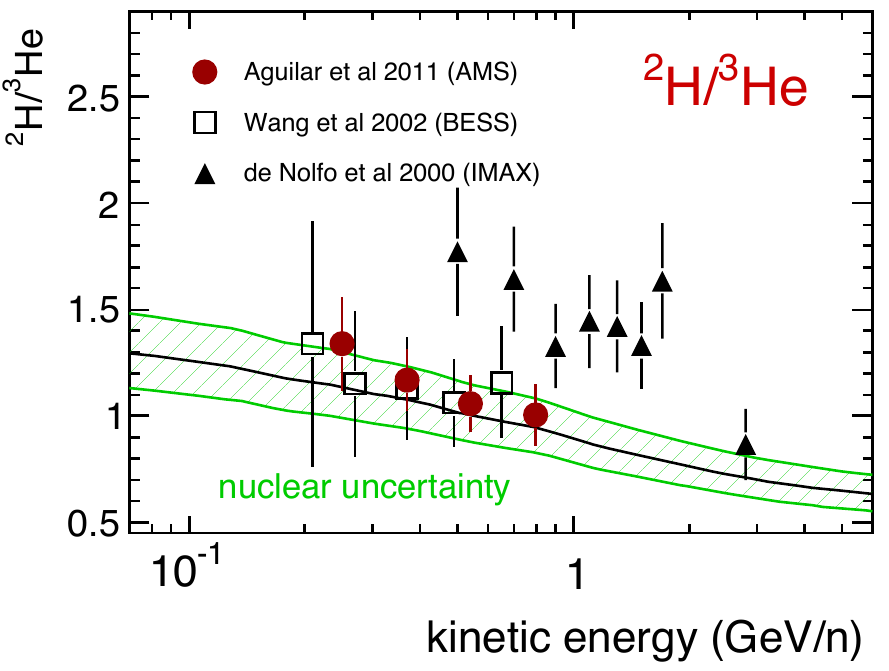}
              \caption{Cosmic ray abundance ratio as a function of the cosmic ray kinetic energy. The nuclear cross sections entering cosmic ray calculations are much more uncertain than the upcoming
               AMS-02 measurements of isotopic ratios, which will be an order of magnitude more precise than the AMS-01 data
               points in this $\ip{H}{2}/\ip{He}{3}$ plot from Tomassetti~\protect\cite{Tomassetti2012}.}
              \label{fig:1}
              \end{figure}
               
              The principle behind indirect detection of particle dark matter is that dark matter particles transform into ordinary
              particles, which are then detected or inferred. Our galaxy is inside a halo of dark matter particles that wander around
              randomly and occasionally annihilate producing otherwise rare cosmic rays, like positrons, antiprotons, and photons
              with special spectra. Many cosmic ray and photon detectors have been searching for these signals from dark matter in
              cosmic ray and gamma-ray fluxes. 
               
              The Alpha Magnetic Spectrometer (AMS), under the direction of Samuel Ting, flew a prototype for 10 days on the Space
              Shuttle Discovery in June 1998 (AMS-01), and has been collecting data on the International Space Station since May
              2011 (AMS-02). The first science results, presented after this conference~\cite{Aguilar2013}, show the positron flux
              measured with the unprecedented precision of a few percent from 500 MeV to 350 GeV. Ten times more data are expected
              and thus a much better precision.
               
              The theoretical models of cosmic ray propagation (GALPROP~\cite{galprop} and DRAGON~\cite{dragon}) are more uncertain
              than the AMS-02 data (see e.g.\ note 17 in Ref.~\cite{Aguilar2013}). In these models, cosmic rays diffuse in a $\sim10\times40$
              kpc region of random magnetic fields surrounding the Galactic disk. Primary cosmic rays (p, $\ip{He}{4}$, C, N, O,
              $\ldots$, Fe, $\ip{Ni}{64}$) are produced in supernova remnants, as first evidenced in Ref.~\cite{Ackerman2013}. Secondary
              cosmic rays ($\ip{H}{2}$, $\ip{He}{3}$, $\ip{Be}{7,9,10}$, $\ip{B}{10,11}$, $\ldots$, $\ip{Al}{26}$, $\ip{Cl}{35}$,
              $\ip{Mn}{54}$, $\ldots$) are produced in cosmic ray collisions with the interstellar medium, which is 90\% H and
              10\% He in mass. The ratio of secondary to primary fluxes carries information on the astrophysical model. AMS-02 is
              expected to measure many of the important isotopic ratios to $\sim1$\% precision up to Fe and $\sim100$ GeV/nucleon,
              and much higher precision at lower energies~\cite{Casaus2003Sapinski2005}.
               
              The nuclear physics implemented in GALPROP is impressive: a nuclear reaction network from $\ip{Ni}{64}$ downward; nuclear
              decays, mostly $\beta$, from the Nuclear Data Sheets; total $\p(\p,x)$ and $A(\p,x)$ inelastic cross sections adapted
              from Ref.~\cite{Tan1983Letaw1983Barashenkov}; total $A(\ip{He}{4},x)$ inelastic cross sections from fits to data; $A(\p,x)B$
              spallation cross sections from LANL-T16, CEM2k and LAQGSM~\cite{Mashnik1998Mashnik2004}, Silberberg-Tsao's
              YIELDX2000~\cite{Silberberg} and/or Webber {\it et al.}~\cite{Webber1990}, with special fits to data for production of
              $\ip{H}{2,3}$, $\ip{He}{3}$, Li, Be, B, Al, Cl, Sc, Ti, V, Mn; $A(\ip{He}{4},x)B$ spallation cross sections from Ref.~\cite{Ferrando1998}. 
               
              Despite this admirable nuclear physics collection, the $A(\p,x)B$ and $A(\alpha,x)B$ spallation cross sections are much more
              uncertain than the upcoming AMS-02 measurements. Tomassetti~\cite{Tomassetti2012} has provided an example of this, using a
              compilation of nuclear cross sections for $\ip{H}{2}$ and $\ip{He}{3}$ production in interstellar space --
              $\ip{He}{4}(\p,{\rm d})\ip{He}{3}$, $\ip{He}{4}(\p,\p\n)\ip{He}{3}$,  $\ip{He}{4}(\p,2\p)\ip{H}{3}$,
              $\ip{He}{4}(\p,\p{\rm d})\ip{H}{2}$, $\ip{He}{4}(\p,\p\p\n)\ip{H}{2}$, $\ip{He}{4}(\p,\p\p\n\n)\ip{H}{1}$,
              $\p(\p,\pi)\ip{H}{2}$ -- and modified parametrizations from Ref.~\cite{Cucinotta1993} (see Fig.~\ref{fig:1}).
               
              Thus the author wishes for better measurements of as many $A(\p,x)B$ and $A(\alpha,x)B$ cross sections as possible
              up to 100 GeV for long-lived ($\lesssim1$ Myr) isotopes with $A\lesssim 64$, which are the progenitors of the H, He,
              Li, Be, B, etc.\ cosmic ray fluxes that will soon be measured to $\sim1$\% precision by AMS-02.
               
              \vspace{0.5\baselineskip}
              \begin{center}
              {\bf III. ~ RADIOACTIVE BACKGROUNDS IN DIRECT WIMP SEARCHES}
              \end{center}
               
              In direct dark matter detection one searches for dark matter particles that arrive on Earth and scatter off nuclei in
              a detector. The only expected signal is some energy deposition, and since almost anything may deposit energy in a
              detector, the name of the game is to operate in low background environments with highly efficient background
              discrimination. This is a very active field and dozens of detectors scattered around the world are taking data or
              will become operational within a year or so. 
               
              \begin{figure}[!t]
              \includegraphics[width=0.9\columnwidth]{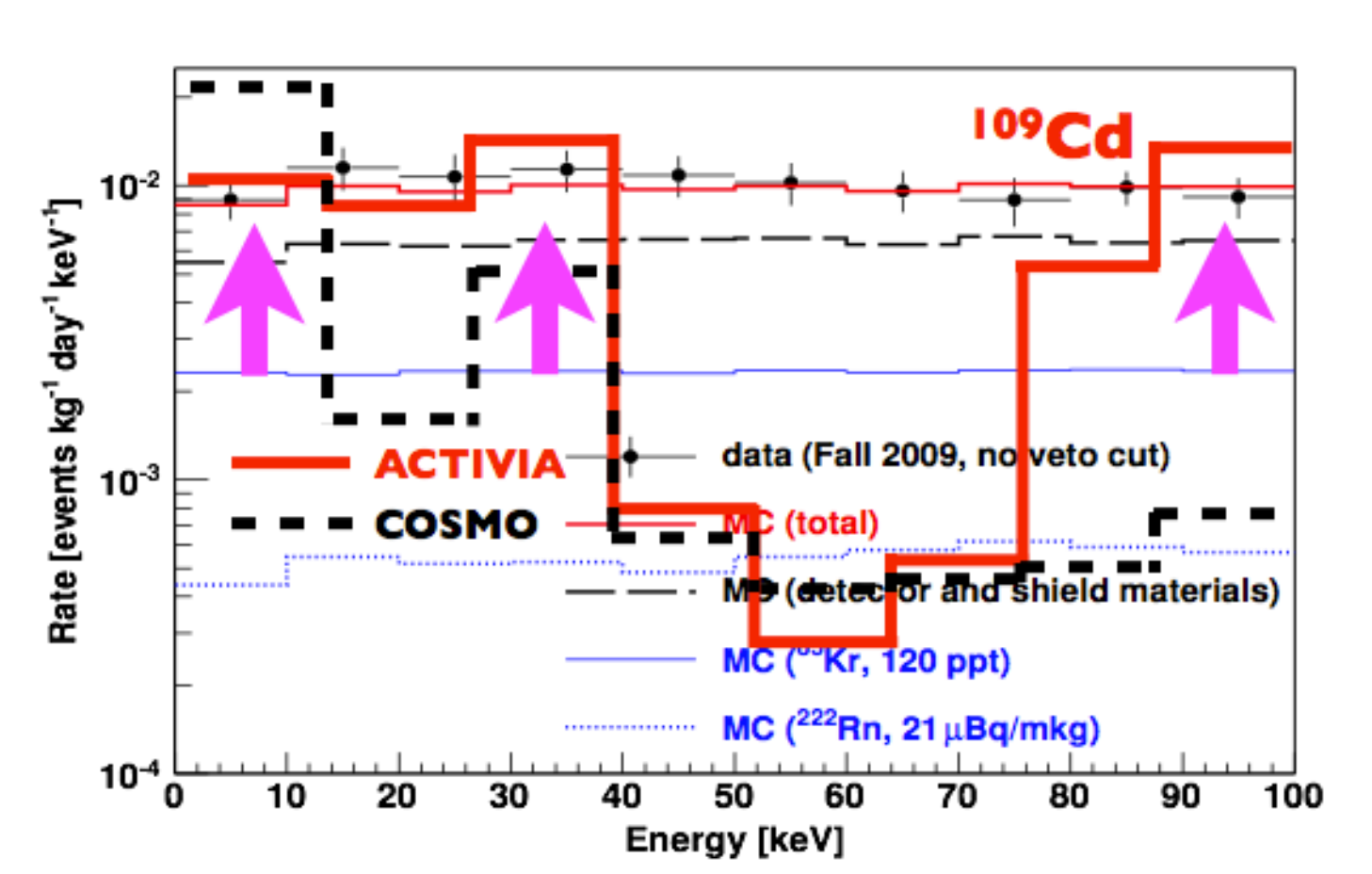}
              \caption{Measured electromagnetic background in XENON100 (black crosses) compared with Monte-Carlo simulations for the cryostat
               radioactivity (thin black and red lines) and the liquid xenon radioactivity (thick red and black lines). The arrows
               point to energy bins in which the Monte-Carlo background exceeds the measured background. (Figure obtained from
               zooming and overlapping figures in~\protect\cite{Aprile2011,Kish2011}.)}
              \label{fig:2}
              \end{figure}
               
              Because of the intrinsic difficulty of distinguishing a neutron background from a dark matter signal, understanding
              the radioactive background in direct detection experiments is very important. The XENON100 study of their
              background~\cite{Aprile2011} is instructive in regard to the nuclear physics involved. Almost all of the XENON100
              radioactivity in the cryostat steel can be accounted for from isotopic measurements of $\ip{Kr}{85}$, $\ip{Rn}{222}$,
              and highly sensitive germanium spectroscopy of similar material~\cite{Laubenstein2009}, adjusted for exposure time and
              extra $\ip{Mn}{54}$ (only a small excess remains unaccounted for around 1 MeV). 
               
\begin{figure}[!t]
              \relax\hbox to 1.2\columnwidth{
              \hspace{-1em}\includegraphics[width=0.51\columnwidth]{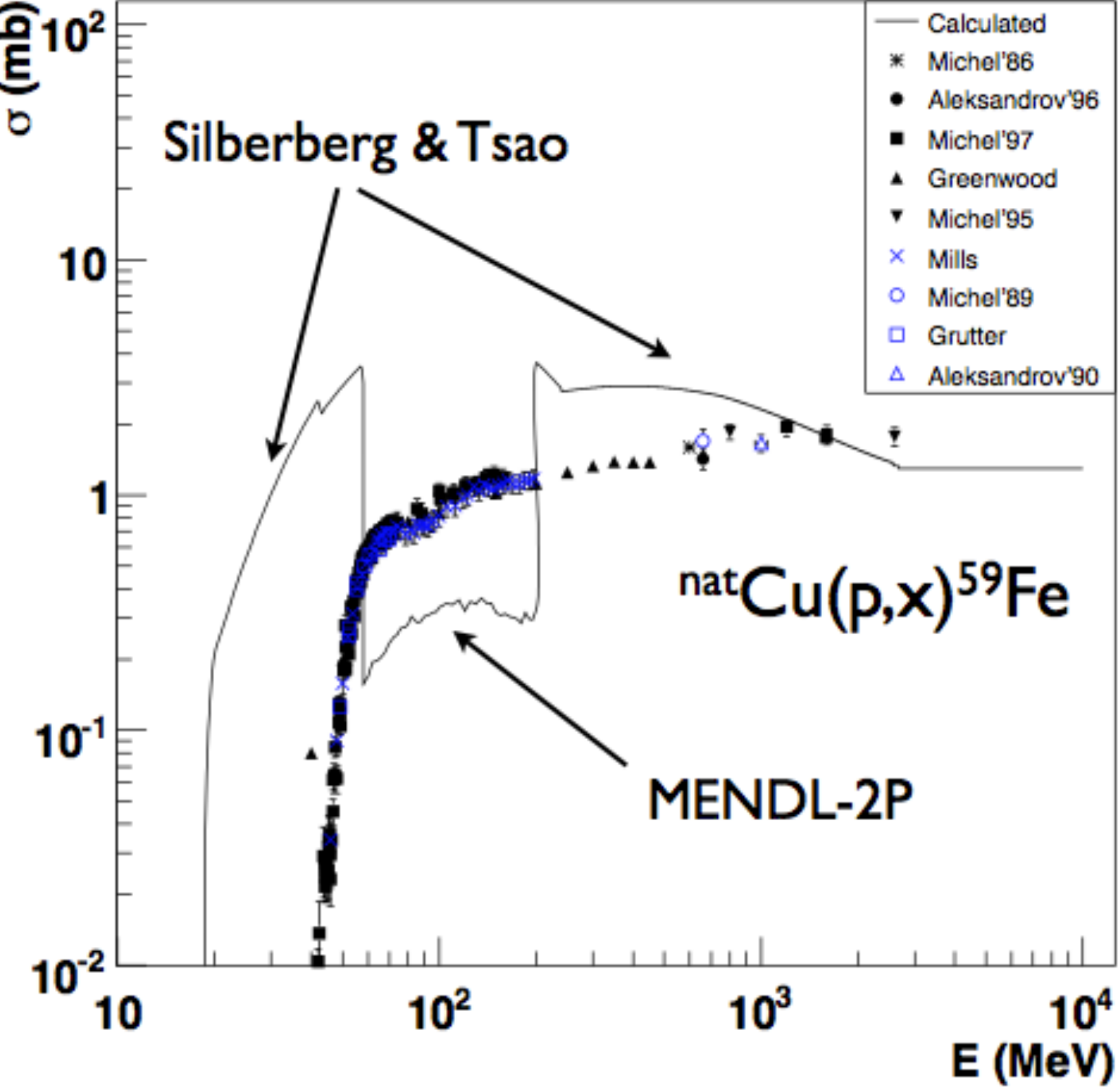}
              \hspace{-0.2em}\includegraphics[width=0.51\columnwidth]{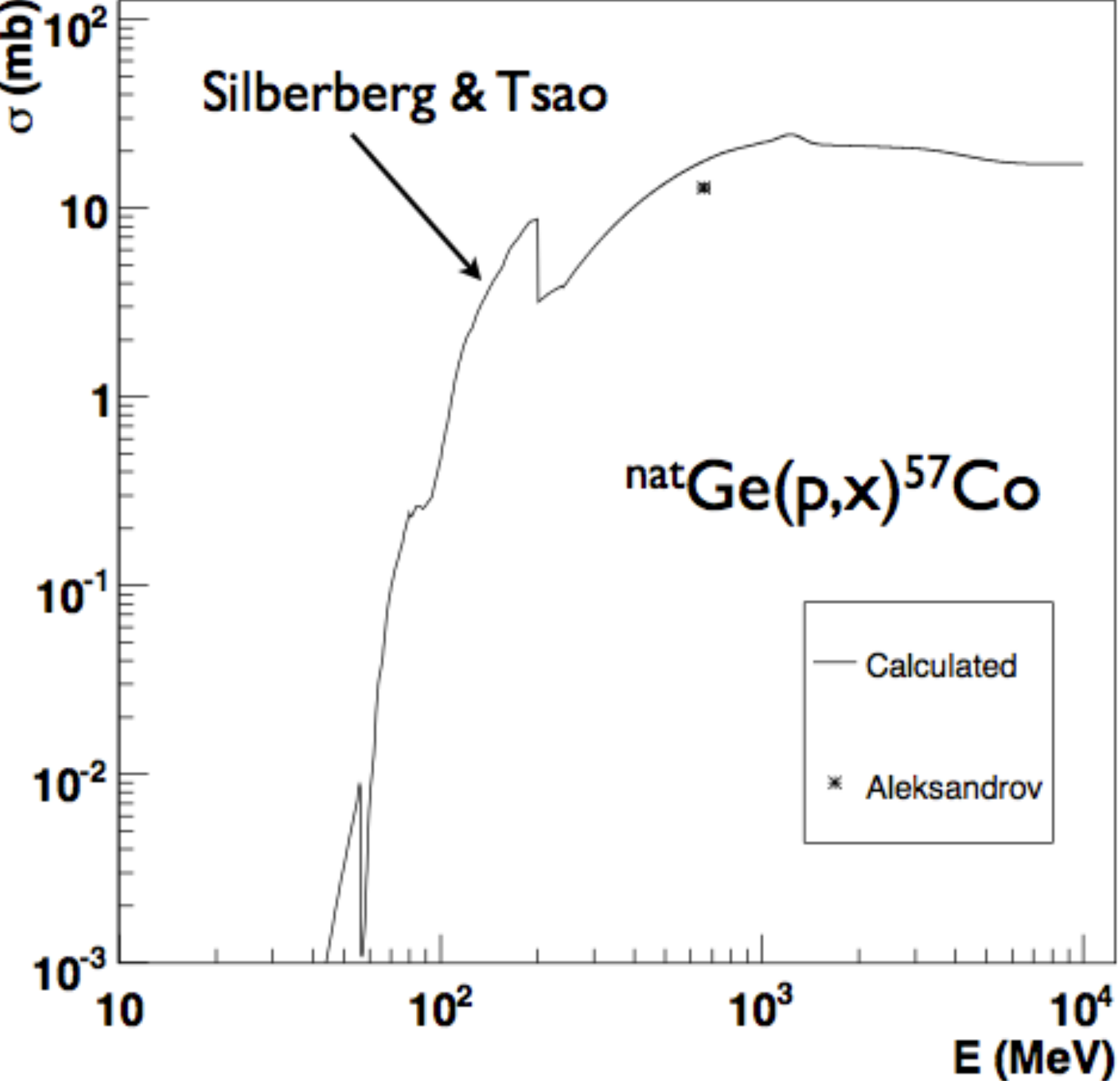}
              \hfill}
              \caption{Examples of activation cross sections used in ACTIVIA (solid line) compared to data (see~\protect\cite{Back2008}
              for larger versions). Can we rely on the cross section in the right panel, which contributes to the background in
              the dark matter region?}
              \label{fig:3}
              \end{figure}

              However, radioactivity from the liquid xenon target itself, which arises from neutron activation of xenon, is poorly
              estimated by existing codes (ACTIVIA~\cite{Back2008}, COSMO~\cite{Martoff1992}, TALYS~\cite{Koning2005}),
              which give results differing by orders of magnitude~\cite{Aprile2011,Kish2011}. In particular, in the region  of
              relevance for dark matter searches ($\lesssim100$ keV ), the simulated background rate exceeds the measured rate
              (Fig.~\ref{fig:2}). A look at some cross sections used in ACTIVIA  (Fig.~\ref{fig:3}) suggests a possible
              explanation: the activation cross sections are too uncertain. 
               
              Therefore the author wishes for more data, or better evaluated data, or better models for $(\n,x)$ in Ge, Xe, Ar, which
              are the target nuclei in large upcoming dark matter experiments such as SuperCDMS, LUX, DarkSide, XENON-1T, EURECA, and DARWIN.
               
              \vspace{0.5\baselineskip}
              \begin{center}
              {\bf IV. ~ SPIN STRUCTURE FUNCTIONS}
              \end{center}
               
              Dark matter scattering off nuclei enters both direct and indirect detection strategies, an example of the latter
              being scattering and capture of dark matter particles into stars. WIMP-nucleus scattering can be either spin-dependent
              or spin-independent, and indeed in many particle physics models, the dark matter particles have non-zero spin and
              interact with the spin of the individual nucleons inside the nucleus. The spin-independent form factor (the Fourier
              transform of the nucleon number density) is relatively well understood theoretically and experimentally, using for
              example the electric charge form factor measured in muon scattering as a proxy. The analogous quantities for the
              spin-dependent part (the spin structure functions) are instead the main uncertainty in the calculation of the
              spin-dependent cross section. 
               
              The spin structure functions quantify the distribution of the nucleon spins inside the nucleus. The spin-dependent
              cross section for the elastic scattering of a spin-$\tfrac{1}{2}$ WIMP $\chi$ of mass $m_\chi$ off a spin-$J$ nucleus
              of mass $M$ with momentum transfer $q$ can be put into the form~\cite{Engel1991Tovey2000}
              \begin{align*}
              \!\!\!\!\!\!\!\!\!\!\!\!\!\!
              \sigma_{\rm SD}(q) = \frac{32\mu^2G_F^2}{2J+1} \left[ a_{\p}^2 S_{\p\p}(q) + a_{\p} a_{\n} S_{\p\n}(q) + a_{\n}^2 S_{\n\n}(q) \right],
              \end{align*}
              where $\mu=m_\chi M/(m_\chi+M)$ is the reduced WIMP-nucleus mass, $G_F$ is Fermi's constant, the $a_{N}$
              ($N=\p,\n$) are effective coupling constants defined so that the four-particle WIMP-nucleon vertex is $2\sqrt{2}G_F a_{N}
              \bsigma_{N} \cdot \bsigma_{\chi} $ (the $\bsigma$'s are the Pauli matrices), and the $S_{NN'}(q)$ are the spin structure functions.
               
              In detail, the WIMP-nucleus spin-spin interaction Hamiltonian is
              \begin{align*}
              H_{\rm spin-spin} = - \int {\bf s}_{\rm DM}({\bf r}) \cdot \Big[ a_0 {\bf s}_0({\bf r}) + a_1 {\bf s}_1({\bf r}) \Big] \, d{\bf r} ,
              \end{align*}
              where ${\bf s}_{\rm DM}({\bf r})$ is the WIMP spin density, and the ${\bf s}_T({\bf r})$ ($T=0,1$) are the proton and
              neutron spin densities in the isospin basis
              \begin{align*}
              {\bf s}_T({\bf r}) = \sum_{i=1}^{A} \, \frac{\bsigma(i)}{2} \, \omega_T(i) \, \delta({\bf r}-{\bf r}_i) .
              \end{align*}
              Here $\omega_0=1$, $\omega_1=\tau_3$ (the third isospin matrix), $a_0=a_{\p}+a_{\n}$, and $a_1=a_{\p}-a_{\n}$. For
              WIMP-nucleus scattering, the matrix elements of the WIMP spin current with initial (final) momentum and spin
              projection ${\bf p} m_s $ (${\bf p}' m'_s $) are $ \langle {\bf p}'m_s' \vert {\bf s}_{\rm DM}({\bf r}) \vert
              {\bf p} m_s \rangle = \langle m_s' \vert {\bf S}_{\rm DM} \vert m_s \rangle e^{i{\bf q}\cdot {\bf r}}$, where
              ${\bf S}_{\rm DM}$ is the WIMP spin operator and ${\bf q}={\bf p}-{\bf p}'$ is the momentum transfer.
              The $e^{i{\bf q}\cdot {\bf r}}$ term gives rise to the spin form factors (Fourier transforms) $\int {\bf s}_T({\bf r})
              e^{i{\bf q}\cdot {\bf r}} d{\bf r}$, which are expanded in multipoles
              \begin{align*}
              \int {\bf s}_T({\bf r}) e^{i{\bf q}\cdot {\bf r}} d{\bf r} = 4  \pi \sum_{\lambda l m} i^{l+\lambda} \,
              {\bf Y}^{(\lambda)}_{lm}(\widehat{\bf q}) \, s^{(T,\lambda)}_{lm}(q) .
              \end{align*}
              Here the ${\bf Y}^{(\lambda)}_{lm}(\widehat{\bf q})$ ($\lambda=0,\pm1$) are transverse-electric, transverse-magnetic,
              and longitudinal vector harmonics.~The spin structure functions then follow as
              \begin{align*}
              S_{TT'}(q) = \sum_{\lambda l} \, \langle J \vert\vert s^{(T,\lambda)}_l(q) \vert\vert J\rangle ^* \,
              \langle J \vert\vert s^{(T',\lambda)}_l(q) \vert\vert J \rangle .
              \end{align*}
               
              Theoretical calculations of spin structure functions are available, assessed by comparison with magnetic moments and
              magnetic dipole transitions~\cite{spinstructurefunctions}. However the author is not aware of any data on the spin
              structure functions. Notice in this regard that the typical momentum transfer in WIMP direct searches is
              $q=\sqrt{2ME_{\rm recoil}} \sim 50$ to 150 MeV/$c$ in I and Xe, $\sim 40$ to 120 MeV/$c$ in Ge, and $\sim 15$ to
              45 MeV/$c$ in F. Notice also that the nucleon spin density is similar but not identical to the axial current density
              appearing in nuclear weak interactions.
               
              Therefore the author wishes for experimental data on the nucleon spin densities (spin structure functions) at
              $\sim 10$ to $\sim100$ MeV/$c$ in nuclei of relevance to direct WIMP searches, such as $\ip{C}{13}$, $\ip{O}{17}$,
              $\ip{F}{19}$, $\ip{Na}{23}$, $\ip{Ca}{43}$, $\ip{Ge}{73}$, $\ip{I}{127}$, $\ip{Xe}{129,131}$, $\ip{Cs}{133}$, $\ip{W}{183}$. 
              The author is unsure about which experimental methods are appropriate to measure the spatial distribution of spin
              (not magnetization, which also contains a contribution from orbital motions) inside these nuclei.
               
              ~\\ The author is grateful to Dr.\ J.-Y. Lee for suggesting this conference and for helping with nuclear physics
              terminology, and to the conference organizers, in particular Dr. A. Sonzogni, for the enthusiasm and interest shown
              in the topic. The author's research is supported in part by the National Science Foundation under Award PHY-1068111.
               
              %%% IMPORTANT: When preparing bibliography observe strictly the layout below (initials in front of the names,
              %%% no names in capitals, not more than four authors, no titles, volume number in bold, year at the end within parenthesis,
              %%% dot (.) at the end.

              \end{document}